\documentstyle[draft]{mn}
\include{epsf}
\def\Ha{\ifmmode {\rm H}\alpha\,\, \else H$\alpha$\,\,\fi}
\def\Hb{\ifmmode {\rm H}\beta\, \else H$\beta$\,\fi}
\def\deg{$^{\circ}$\,\/}
\def\etal{{\it et al. \/}}
\def\ukirt{{\it UKIRT\,\,\/}}
\def\vlbi{{\it VLBI\,\/}}
\def\cgs4{{\it CGS4\,\/}}
\def\ergcms{erg~cm$^{-2}$~s$^{-1}$}
\def\kms{\ifmmode {\rm km\ s}^{-1} \else km s$^{-1}$\fi}
\def\qo{\ifmmode q_{\rm o} \else $q_{\rm o}$\fi}
\def\Ho{\ifmmode H_{\rm o} \else $H_{\rm o}$\fi}
\def\Hubble{km s$^{-1}$ Mpc$^{-1}$}
\def\lha{\ifmmode L_{\Ha} \else $L_{\Ha}$\fi}
\def\lo{\ifmmode L_{\rm o} \else $L_{\rm o}$\fi}
\def\R{\ifmmode R\, \else $R$\,\fi}
\def\m{\ifmmode M\, \else $M$\,\fi}
\def\ba{\ifmmode \beta_{\rm a}\, \else $\beta_{\rm 
a}\,$\fi}
\def\b{\ifmmode \beta\, \else $\beta\,$\fi}
\def\g{\ifmmode\gamma\else$\gamma$\fi}
\def\d{\ifmmode \delta \else $\delta$ \fi}
\def\th{\ifmmode\theta\else$\theta$\fi}
\title[A Disc in the SL Quasars?]
{ Is There a Disc in the Superluminal Quasars? }
\author[E. Rokaki, A. Lawrence, F. Economou \& A. 
Mastichiadis]
{E. Rokaki,$^{1,2,3}$ A. Lawrence,$^{2,3}$ F. 
Economou$^{3,4}$ \& A. Mastichiadis$^{1}$\\
$^1$ Section of Astrophysics, Astronomy \& Mechanics, 
University of
Athens, 15784 Zografos, Athens, Greece\\ 
$^2$ Institute for Astronomy, University of Edinburgh,
Royal Observatory, Blackford Hill, Edinburgh, UK\\ 
$^3$ Physics Department, Queen Mary and Westfield College, 
Mile End 
Road London E1 4NS,UK\\ $^4$ \ukirt, JACH, Hawaii} 
\date{} 
\def\LaTeX{L\kern-.36em\raise.3ex\hbox{a}\kern-.15em
\kern-.1667em\lower.7ex\hbox{E}\kern-.125emX}

\begin{document}
\label{firstpage}
\maketitle
%
%
\begin{abstract}
We look for the expected signature of an accretion disc by 
examining the properties of the \Ha emission line versus 
viewing angle in a sample of 22 superluminal (SL) quasars. 
The Doppler factor \d, jet velocity \g , and viewing angle 
towards the jet \th , are derived from published radio and 
X-ray data.  
Most of the \Ha spectra (fourteen) have been observed at the 
United Kingdom Infrared Telescope (\ukirt) and are reported 
here. About one fourth of the SL objects have weak or 
absent \Ha emission lines, with small equivalent widths 
(EW). These have high optical 
polarization, radio core dominance, Doppler factor and most 
of them high apparent SL velocity and low viewing angles. 
Therefore these weak-EW objects almost certainly have 
relativistically beamed 
optical continua. The strong-EW objects also show a 
clear beaming effect, but a much weaker one, with line EW 
varying by only a factor three while radio core dominance 
varies by a factor of several hundred. The correlation of 
EW with \th\ is quantitatively in good agreement with 
the prediction of a flat accretion disc with limb 
darkening. The weak- and strong-EW sources also show an 
anti-correlation of line velocity width with the various 
beaming indicators. Again, the correlation with the derived 
viewing angle \th\ shows a quantitative agreement 
with the effect expected for an axisymmetric structure with 
velocity dominated by rotation. The line emission cannot 
come from the surface of the disc, or the line beaming 
would cancel the continuum beaming. However it could come 
from an axisymmetric system of clouds co-rotating with the 
accretion disc. 
\end{abstract}   
%
%

\section{INTRODUCTION}
Attempts to incorporate all AGN (active galactic nuclei) 
activity within some framework (``Unified Schemes") have
focused on the importance of the orientation of the observer 
with respect to the source, in three separate ways. (i) 
The most dramatic effect is relativistic Doppler-beaming in 
a jet of radio-emitting plasma aligned close to the line of sight. 
Amplification of the apparent brightness of the source and apparent 
superluminal motions are then expected (Rees 1966; Lind \& 
Blandford 1985; Ghisellini \etal\ 1993; Urry \& Padovani 1995). 
This effect explains the existence of ``blazars'', objects 
dominated by a polarised and highly variable continuum, as 
well as one sided radio jets and other phenomena. This is 
the most well established orientation effect, but only applies 
to the minority of radio-loud objects. (ii) The second effect is 
that of absorption or obscuration, most probably by an 
equatorial disc or torus, invoked to account for the 
dissimilarities between type~1 and type~2 Seyfert galaxies 
(Lawrence \& Elvis 1982, Antonucci \& Miller  1985, Antonucci 1993).
(iii)  Finally a third orientation effect is associated with an 
accretion disc geometry: (a) a simple optically thick, geometrically thin 
disc viewed inclined will appear fainter than when it is viewed 
pole on. However relativistic effects in the vicinity of the 
accreting black hole make edge-on discs brighter (Cunningham 1975). 
(b) If the emission lines share the same geometry they will have 
a broad  velocity width at low latitude. 
(c) The polarization of the disc radiation will increase
with the inclination (e.g. Rees 1975).

Some authors have searched for evidence of this third class 
of orientation dependent effects, generally by using the ratio 
of core radio luminosity to extended radio luminosity (e.g. Orr 
\& Browne 1982; Wills \& Browne 1986; Browne \& Murphy 1987; 
Jackson \etal\ 1989; Jackson \& Browne 1991b; 
Corbin 1997). This ratio is an orientation indicator 
(see for instance Ghisellini \etal 1993, Wills \& Brotherton 1995) 
and these studies have been consistent with the widespread presence 
of optical beaming in radio loud quasars in general 
(as opposed to simply in blazars, e.g. Browne \& Murphy 1987) 
 and a flattened geometry for the emission lines 
(e.g. Wills \& Browne 1986, Corbin 1997). 
The amount of optical beaming seen is far less than seen in 
the radio regime, indicating that it might be due to 
to relativistic beamed continuum seen 
in low contrast against less anisotropic disc emission 
(e.g. Browne \& Murphy 1987; Wills 1991; Wills \etal 1992).
The idea of this paper is to use {\em superluminal sources} 
for which we can derive an actual viewing angle rather than 
an indirect orientation indicator, and so directly test for the 
{\em quantitative} effects of accretion disc surface 
brightness and emission line velocity projection 
(see also Wills \& Brotherton 1995).

In this paper, we study the \Ha emission of a sample of 22 
superluminal sources. This line was chosen because 
it is the strongest low ionization emission line 
(LIL, e.g. hydrogen Balmer, Mg II etc.) for which, there are both 
observational (e.g. Wills \& Browne 1986) and theoretical 
(e.g. Collin-Souffrin 1987) suggestions that its velocity field 
is anisotropic, likely in a rotational disc structure. 
In addition, \Ha line is less affected by absorption or blended 
optical Fe~II multiplet emission than \Hb. Line width is used 
to study velocity projection, and equivalent width to indicate 
the brightness of the underlying continuum. The viewing angle 
is derived from published SL motions, radio and X-ray fluxes, 
and angular sizes, using the method of Ghisellini \etal\ 
(1993). 
For eight sources we use published \Ha data, but for the 
majority we have obtained new data. Most superluminal 
sources are at moderately large redshifts, so this has 
required obtaining IR spectra, which we report in this 
paper.

In Section 2 we summarise the technique used to derive 
viewing angle. In Section 3, we describe the SL sample 
giving relevant data and the parameters derived by the 
Inverse Compton model. In Section 4, we present the new 
\ukirt observations, with detailed source notes in 
Section 5. In Section 6 we present the statistical results 
which are discussed in Section 7 and summarised in Section 8.
The cosmological model assumed in this paper
is \Ho = 75 \Hubble\, and \qo=0.5.     
%
%
\section{METHOD of DERIVING VIEWING ANGLE}
Derivation of the viewing angle relies on two effects - 
superluminal motion, and Doppler boosting.
We follow the method of Ghisellini \etal\ (1993) which we 
now summarise briefly.

Apparent Super-Luminal (SL) motions of individual radio 
structures 
(measured with
\vlbi\, experiments, see for instance Zensus \& Pearson 
1987) have been observed in extragalactic radio sources
for more than 20 years. These motions are usually modelled 
by one or more radiating ``blobs"  moving at relativistic 
velocity  $\beta$ (in units of c) away from a 
stationary ``core". If the viewing angle is \th, the 
apparent transverse velocity is
 
\begin{equation}
\ba = { \b \sin \th \over 1-\b \cos \th} 
\end{equation}

and can exceed the true speed and even 
that of light if the direction of the motion is close to 
the line of sight. The observed SL velocity 
therefore gives a combination of the Lorentz factor \g, 
and jet viewing angle \th\,. The radiating material is 
brightened by the Doppler beaming factor \d , which also 
provides a combination of \g\, and \th\, (\d $=[\g(1-\b 
cos\th)]^{-1}$). If \d can be separately estimated, then, 
following Appendix B of Ghisellini \etal\ (1993),  
\g\ and \th\ are given by

\begin{equation}
 \g = \frac{\ba^2 + \d^2 + 1}{2\d}
\end{equation}
and
\begin{equation}
 \tan{\th} = \frac{2\ba}{\ba^2+\d^2-1} 
\end{equation}

The value of \d is estimated from observations of the flat 
spectrum radio core, assumed to be made up of multiple 
self-absorbed components, by requiring that the Inverse 
Compton flux of the component peaking at the observed radio 
frequency is less than the observed X-ray flux (see 
equation 1 of Ghisellini \etal ). This also requires the 
angular size of the source, from \vlbi\ measurements, 
and an assumption for the underlying synchrotron spectral slope, 
taken to be $\alpha=0.75$. 

We summarise uncertainties in the estimate of \th . (i) The 
value of \d estimated from the X-ray flux is a lower limit, 
and so the derived \th\ is an upper limit. However the value 
is very insensitive to this effect - \d $\propto F_x^{-0.22}$. So 
if the true $F_x$ is actually lower by a factor 10, 100, or 
1000, the value of \d is larger by a factor 1.5, 2.3, and 
3.5 respectively. This factor might increase by up to 30 \% if
the source is not spherical as assumed, but a continuous jet (see 
equation 2 of Ghisellini \etal)  (ii) Radio flux and angular size
errors may also introduce uncertainties on \th, but in most cases
these are not large. (iii) Another possibility is that the 
superluminal velocities represent a pattern velocity which 
is different than the bulk velocity (Lind \& Blandford 1985; 
Vermeulen \& Cohen 1994). (iv) 
Finally the observed value of \ba and thus \th\,
depends on the assumed Hubble constant, but we take that to 
be understood (Freedman \etal\ 2001). If \Ho\, is smaller than 
the assumed value, \th\, is smaller in most cases.
In section 6 we provide 
arguments that \d cannot be wrong by a very large factor, 
but also illustrate the effect on derived values of \th\ of 
assuming that \d is wrong by a factor of a few.

%
%

\section{THE SAMPLE}

The sample of the SL sources was 
selected from the list of Vermeulen \& Cohen, 1994 (VC94). 
The final sample used in this paper is the superset of 
those for which we were able to obtain good quality IR 
spectra at \ukirt, plus those with published (optical) data 
on \Ha. This sample represents more than the one third of 
the known SL extragalactic sources. It is listed in Table 
1. Columns (1), (2), (4), (5), and 
(6) list names, redshift, the fastest measured apparent 
velocity, and the core-to-lobe flux ratio $R$ at 5 GHz as 
given in VC94. Column (3) divides 
the 
objects into three classes depending on radio core 
dominance and optical polarization. Core dominated sources 
with low polarization are shown as ``CD'' and those with 
high polarization ($P > 3\%$) as ``CDP''. Lobe dominated 
sources are shown as ``LD''.  
The limit between CD and LD was taken as log \R =--0.02.  

Columns (7)-(10) give the observed radio core flux and 
angular size ($\theta_d = \sqrt{ab}$ where $a$ and $b$ are 
the major and minor axes in mas) at a given frequency, 
together with the reference. Finally the two last columns 
give the observed X-ray flux at 1 keV together with the 
reference. The radio and X-ray data were then used to 
estimate \d, and thus \th , and \g , as explained in the 
section above. For 3C~395 we have not found any X-ray measurement
published, and we have used optical data instead. 
Table 2 shows the result, with sources 
listed in order of 
increasing \th , which can be seen to correlate clearly 
with 
\d but not with \g, as expected in the beaming model.

%
%

\section{{\bf {\it UKIRT }}\, spectra}

\subsection{Observations and Data reduction}

From the VC94 sample only objects between RA 05$^h$ and 
19$^h$  and Dec. lower than 60$^\circ$ were observed, due 
respectively to the time of year of the observations and 
the declination limit of the 
telescope. Finally we selected sources at redshifts that 
would place \Ha well inside one of the $J, H$, or $K$ 
atmospheric windows. This gave 14 sources which are listed 
in Table~3. 3C 286 (not known as superluminal) was also 
observed during an RA gap in the run and is likewise 
listed in Table 3.
 
The spectra were obtained with the common-user Cooled 
Grating Spectrometer 
CGS4 at the f/36 focus of the 3.9-m \ukirt on Mauna Kea, on 
three consecutive photometric nights starting on UT date 
1994 March 27. The instrument was equipped at this time 
with a 62$\times$58 InSb array detector at the focal plane 
of a 150-mm camera with a slit of width equal to 3 arcsec, 
stepped perpendicular to the dispersion. The  detector 
was stepped in increments of one-third of a resolution 
element in order to properly sample the instrumental 
profile. Different gratings and orders used, depending on 
the  redshift of the 
source, and exposure times, are given in Table~3. 

For most of the $J-$ and $H-$ band observations, a 75 line 
mm$^{-1}$  grating in second order was used. This gave a 
resolving power of 383 and 
0.2 $\mu$m spectral coverage. For one case in the $J-$ 
band, the 150 line 
mm$^{-1}$ grating was used in third order, giving a 
resolving power of 
1000 and a wavelength coverage 0.1 $\mu$m. For observations 
in the 
$K-$band the 75 mm$^{-1}$ grating was used in first order, 
giving a resolving power of 337 and 0.4 $\mu$m spectral 
coverage.

Standard nodding procedures were followed to subtract the 
sky background, with the slit oriented east-west on the sky. 
While observing, we nodded on to the detector 
so that we could integrate on the object while obtaining frames 
for sky subtraction. Bias frames were subtracted and the 
observation images were then flat-fielded. An argon lamp and 
vacuum wavelengths 
are used here for the wavelength calibration. Correction for 
atmospheric extinction was applied by observing a spectral 
standard star at a similar airmass to the object. Flux 
calibration was achieved by extrapolating $J$, $H$ and $K$ 
magnitudes from observed $V$ magnitudes of flux standards 
according to known spectral type using the model of 
Koorneef (1983). 

We note that the measured counts in 
the ``sky'' frames were not always consistent with those in the 
``object'' frames. From examination of the photometric consistency 
of the data we are led to conclude that any flux measurements have 
an uncertainty at the level of 10\% (except for 3C 380 in which 
the uncertainty is $\sim $20\%). Data reduction was carried out 
using the CGS4DR package at the QMW and Edinburgh Starlink nodes. 

Figure~1 shows the observed spectra. The 14 SL sources are 
presented in RA order. The spectra of 3C~286 and the three 
serendipitous sources are given at the end. Note that there 
are only three SL sources where \Ha was not detected. These 
are 3C~216, 4C~29.45, and 4C~56.27. In a further two 
sources, \Ha is detected but is rather weak (3C~279 and 
B2~1308+32). In the remaining 9 a strong line was 
detected. Together with the 8 additional published sources 
that we use, this produces three sub-samples that we refer 
to : SL sources ( 22 objects), a sample of 19 SL 
sources with detected \Ha emission, and a sample of 16 
strong-EW sources.


\subsection{Emission line parameters }

Emission line measurements were performed using the {\it 
DIPSO} package on the ROE Starlink node. In Table 4 we 
present the observed \Ha flux, rest frame EW 
and FWHM of the flux calibrated spectra. From these data 
both line and continuum luminosities ($L_{H\alpha}$ and 
$L_o$) can be derived.  The largest source of error comes 
from uncertainties in setting the local continuum (fitted 
by a first order polynomial). To 
minimize these errors each line was measured on at least 
three different occasions. 
In most cases the uncertainties are less than 10\%. 
Experimentation with deblending the narrow lines (\Ha, 
[NII], [SII]) suggests that 
in most objects its presence does not introduce more than a 
3\% error  into the flux of \Ha. This maybe due to 
our inability to resolve and exclude the narrow lines 
from either sides of \Ha. When the narrow \Ha\, emission 
line is visible however, its contribution can change substantially 
FWHM. In these cases the measurements are corrected from narrow
line contribution.
 
In addition to our new data, Table 5 gives the \Ha emission line 
parameters of 8 SL sources with published \Ha spectra. These data
are less homogeneous than the \ukirt spectra. We have searched for 
total line fluxes and EW (including also NLR), but corrected FWHM 
(from NLR) values (references are given in Table 5).

%
%

\section{Comments on individual objects}

DA 193. The measured line flux and rest EW of the \Hb\,
 emission line
in this source are 1.7 10$^{-14}$ \ergcms and 40 \AA\, 
respectively. 
The total [OIII] flux is estimated as 9\,10$^{-15}$ 
\ergcms. 
The two spectra obtained of this source show an increase 
of the 
continuum flux at shorter wavelengths.

4C 39.25. In the field of 4C 39.25 we have detected what 
appears to 
be an extended object (it occupies two detector rows) about 
12 arcsec 
west of our source. We can identify this in the Schmidt 
plate as a point 
source which is brighter in the blue. From this and the 
spectrum (shown 
in Fig.~1) we conclude that it is a high redshift galaxy. 

3C 279. The broad feature at 1.1 $\mu$ of the spectrum is 
most likely 
due to the non-fully corrected absorption feature of the 
spectral standard.

3C 334. The high resolution spectrum obtained in this 
source (see Fig 1) shows a structured \Ha emission line profile 
but it has also a low signal-to-noise ratio. It is also likely 
not resolved in the continuum, and is arbritarily multiplied 
in flux for comparison with the lower resolution \Ha spectrum.

3C 395. Two other objects, most likely stars,  were 
detected in the slit at 3'' east and at 15'' west 
of the source. Their spectra are shown in Fig.~1.

%
%

\section{Statistical Results }

We have applied the Spearman rank correlation test (Press 
\etal\ 1988) to a set of nine parameters - to three 
emission line parameters (FWHM, EW, and \lha ), to the 
optical luminosity at the \Ha wavelength (\lo ), to three 
orientation indicators ($R$, \ba\, and \d\,), to \th , and 
finally, to the Lorentz factor, \g . The resulting matrix 
of the correlation coefficient $r_s$, and the two sided 
significance level of its deviation from zero, $p$, is 
shown in Table 6. The test was applied both to the 
sub-sample with detected \Ha\ lines (N=19) and to the 
subsample with strong lines, divided at EW$\geq 158\,\AA$ 
(N=16).  

The strongest trend is seen between the \Ha\ line 
and continuum luminosity (\lha\ and \lo\ respectively), 
especially when only the strong-EW sources are 
considered. This is however a selection effect, the result
of plotting two quantities both multiplied by the same function
of redshift. This trend is shown in Fig 2a where the three 
classes of objects CDP, CD, and LD are shown with separate 
symbols. 
(The three objects with \Ha\ upper limits are also 
shown here). The strong- and weak-EW objects are 
very clearly distinct (see Fig 2b). The weak-EW objects have an 
EW at least a factor five less than the weakest 
strong-EW object. Note that the range of luminosities 
\lha\ is not very different in the strong- and 
weak-EW objects. As we shall see below, the FWHM 
does not show any such clear discontinuity, so that the two 
distinct trends in Fig 2 is almost certainly due to a 
difference in the continuum, not in the line. It is also 
striking that all the weak-EW objects are polarised. It 
seems then extremely likely that the weak-EW objects 
have a continuum enhanced by relativistic beaming. Very 
likely there are two continuum components, the first of 
which is weak in the strong-EW objects, but is boosted 
to dominance in the weak-EW objects. The object 3C~345 
with the largest EW amongst the CDP objects, looks like a good 
intermediate case, with significant but not 
dominant relativistic beaming. Broad band polarization and 
flux density spectra for this object do show the presence
of two emission components (Wills 1991): one polarized, 
likely synchrotron jet radiation, and a second unpolarized possibly  
from a disc. Optical and Near-IR polarimetric and photometric
monitoring also support the two continuum emission components in this source
(Smith \etal 1986).

In all our subsequent 
plots, we reclassify objects as simply strong-EW or weak-EW. 
The weak-EW class is identical with the CDP class except 
for 3C~345, which has been classed as strong-EW. Within the 
strong-EW objects, we maintain a distinction between LD 
and CD objects.

Equivalent width EW is correlated with the 
various orientation indicators, as indicated in Table 6 and 
Fig. 3, which collects these correlations graphically. (Note 
the inverted abscissa for the graphs versus \d , $R$, and 
\ba , chosen so that these can be easily compared with the 
correlation with \th .) Amongst the strong-EW objects, 
the strongest 
correlations are with \d\ and with \th , in that order. The 
excellent correlation with \d\ is as expected, as this 
indicates the amount of Doppler beaming. The fact that it 
is significantly better, with much less scatter, than the 
correlation with the indirect orientation indicator \R, 
 indicates that despite the 
uncertainties, stressed in Section 2, the derived values of 
\d\ must be quite close to the correct values. The 
correlation with \R\, is striking in a special respect
as \R\, is a kind of direct analogue of EW. 
 While EW changes by only a factor of a few, \R\, 
changes by a factor of several hundred. There is an anisotropy
connected with the radio beaming, but it cannot be the same 
jet beaming effect (at least for the LD objects), but rather 
 a second ``weak beaming'' effect. There is a 
very strong positive 
correlation between the line EW and the outflow angle of 
the radio jet in the whole sample, but this is dominated by 
the cliff at small angles where the jet beaming sets in. 
The trend slightly weakens in the strong-EW sources and 
indicates grossly
an increase of the EW by a factor of $\sim$3 when the 
viewing angle 
turns from  5 to 40 degrees.

Next we look at the correlations of FWHM, which are 
collected in Fig 4. FWHM is correlated with all the 
orientation indicators, indicating a decrease of the 
emission line velocity field 
with increasing beaming. The direct correlation with \th\ 
is suggestive of axisymmetric motions for the BLR. The most 
significant correlation seen is with viewing angle \th , as 
shown in Table 6.  This is an indication that in 
deriving \th\ we have done something meaningful. As further 
support for the reality of the \th\ correlation, note that 
Table 6 shows no correlation between FWHM and the jet 
Lorentz factor \g . The weak-EW objects fit well into 
the same trend as the strong-EW objects, being simply at 
one end of the trend, namely at small viewing angle. The 
trends weaken somewhat when considering the strong-EW 
sample, which is however in a reduced range 
of \th.

%
%

\section{Discussion}

Our main aim has been to compare \Ha emission properties 
with observer viewing angle. The most exciting result is 
that both velocity and equivalent width do indeed correlate 
with orientation. The actual \th\ derived from the SL 
motion plus Inverse Compton limit seems more reliable than 
the various indirect orientation indicators, as it gives 
stronger trends, and is physical. 

The FWHM vs \th\ correlation is suggestive of a flattened 
structure for the line emitting material. Fig 5a tests this 
quantitatively, plotting $V=V_0 \sin \th $ with 
 normalised line velocity of 
$V_0=13,000$ km s$^{-1}$ at \th =90\deg. This is the disc fit of Wills
\& Browne 1986 to the \Hb\,  in a larger AGN sample, extended to 
larger jet viewing angles. The figure also shows 
the effect of an isotropic component with $V=2,000$ 
km s$^{-1}$ (see for instance Brotherton \etal 1994, Corbin 1997) 
added in quadrature to the axisymmetric component.
The agreement is good. The scatter is 
considerable, but we should certainly of course expect some 
dispersion in $V_0$.  How sensitive is this agreement with 
the predictions of an axisymmetric structure to the 
possible errors in \th ? As described in section 2, the 
value of \th\ is strictly an upper limit, as the derived 
value of \d is a lower limit. However, as argued there, \d 
is very insensitive to the true Inverse Compton flux, and 
the correlations discussed in the previous section confirm 
that \d is unlikely to be in error by more than a factor of a 
few. We illustrate this sensitivity in Fig 6a by deriving 
the \th\ values for each source with \d increased by a 
factor three. It can be seen that the effect is still clear 
but the quantitative fit is not as good. It is also 
possible that the bulk speed is smaller than 
the pattern speed responsible for observed SL velocity. 
We do not illustrate the effect of this explicitly, 
but note that the net effect is similar to the \d effect 
shown here.      

An important implication of a flattened structure for the broad line 
emitting region (BLR) concerns the mass \m, of the black hole when this
is determined from the line velocity width assuming gravitational dynamics
and isotropically-oriented motions (e.g. Wandel, Peterson \& Malkan 1999, 
Kaspi \etal 2000).
In a thin Keplerian BLR of inclination angle equal to \th\, the mass 
fit of a given line width is $\propto (\sin^2 \th\sqrt{\cos \th})^{-1}$
which is $\sim 1/\sin^2 \th$ at low \th\ (Rokaki \& Boisson, 1999). 
The inclination therefore may cause a systematic underestimate of \m, 
theoretically up to infinity. In more realistic orbital shape of the
BLR, involving an aspect ratio of its flattening, the inclination
may cause a systematic underestimate of mass by a factor up to 
$\sim 100$ (Krolik 2001). It is statistically implausible 
that the mass of the quasars is underestimated by such a large factor
(this would happen if all sources are observed with \th=0) but it could 
be important for sources found accreting at super-Eddington rates 
(see Collin \etal 2002). The FWHM vs \th\ correlation therefore, 
supports estimates of the black hole mass in which an orientation 
dependence of the line width is applied (e.g. Lacy \etal 2001, McLury \& Dunlop 2002). 

The EW correlation shows the presence of two separate 
beaming effects - a strong effect presumably connected with 
relativistic beaming in the jet, and a second weaker 
effect, producing a factor $\sim 3$ change for \th\ range 
from 5 to 40 degrees. It is tempting to identify 
this second effect with the $cos \theta$ surface brightness 
effect expected from a flat accretion disc. Fig 5b examines 
this possibility quantitatively. The dot-dashed curve shows 
the expected effect of Doppler enhancement for a jet with 
$\g = 30$ (imposed mainly from the objects with large \ba). 
This can explain the difference between CDP 
objects and the rest, but not the general trend. The dotted 
line combines an isotropic continuum with a second separate 
jet-beamed continuum. This also does not fit the overall 
pattern. The dashed line shows the predicted effect for a 
flat accretion disc, with standard limb darkening, where we 
should find $EW = EW_0 [1/3 \cos \th(1+2 \cos\th)]^{-1}$. Fig 
5b plots this with $EW_0=300\, \AA$. Note that apart from the 
normalisation there is no further degree of freedom. The 
solid line combines this disc-beamed continuum with a 
second separate jet-beamed continuum. The overall fit is 
good.

In Fig 6b, we indicate the effect on \th\ of changing 
\d  by a factor of three, as 
explained in the FWHM discussion. The qualitative trend is 
still clear, but the quantitative fit is not as good. Given 
this uncertainty, we cannot hope, for example, to distinguish 
classes of accretion disc models such as flat discs,  
thick discs and flared discs. It is possible to restrict
the range of \th\, in the sample ( e.g. by decreasing \Ho), 
more than in Fig 6b. In this case a jet model with a
broad distribution of \g\, could possibly account for 
the observed correlation, but requiring small \g\, ($<$5) 
for lobe dominated objects (not seen in Table 2).
 
Another worry is that we have implicitly assumed that the 
continuum light has a 
surface brightness effect with \th , but that the line 
emission does not. If the BLR is however axisymmetric, as 
the FWHM effect seems to show, then \Ha emission may come 
from the surface of the disc, and suffer 
a very similar effect with \th , removing the EW vs \th\ 
effect. The effect would be the same as the dotted line in 
Fig 5b, showing isotropic continuum, which clearly does not 
fit the data. However, a likely scenario is that the line 
emitting material is in an axisymmetric structure but is 
{\em not} from the surface of 
an optically thick disc, being rather in a flattened system 
of clouds. Even if the lines are emitted from the surface 
of a disc, they will not necessarily suffer the same limb 
darkening law. 

If the observed optical continuum of the strong line class
is emitted from a disc, it is not likely that it is markedly 
influenced by relativistic effects such as the
gravitational redshift and lensing. These effects are stronger
when the disc is viewed in the equatorial plane (and in the 
Kerr metric, Cunningham 1975), which is not the case for the 
SL objects.

%
%
 
\section{Conclusions} 

In this paper we have investigated the relation of the \Ha 
emission
line with radio beaming and outflow angle in a sample of SL 
sources.
Most of the \Ha spectra were observed by us at the \ukirt 
and are reported here. 

The sample is divided into two classes: 6 weak-EW
and 16 strong-EW quasars. The weak-EW class must have a 
relativistically 
beamed continuum, since it has also high radio core 
dominance and high optical polarization. The strong-EW 
class shows a clear beaming effect, but one that cannot 
be relativistic beaming alone, since the
line EW changes only by a factor of a few. The 
correlation with the actual viewing angle derived from the 
superluminal motions is very clear and nicely consistent 
with this weak beaming being due to the surface brightness 
effect expected from tilting a flat accretion disc. We also 
find that the  FWHM of \Ha is correlated with the
viewing angle in a manner quantitatively consistent with 
the idea that the line emitting material is an axisymmetric 
rotating structure. However, the line emission cannot come 
from the surface of an optically thick disc, otherwise 
the line beaming would cancel the continuum beaming. 
Overall it seems most likely that the optical continuum 
comes from a flat accretion disc whereas the line emission 
is from an axisymmetric system of clouds.

%
%
%

\section*{Acknowledgments}
We are grateful to the \ukirt\, team for their help in the 
observations. We thank Suzy Collin for helpful discussion, and 
Anne-Marie Dumont, Catherine Boisson and Maria Kontizas for help 
in this work. We also thank the referee, Bev Wills, for critical 
comments which improved this work. ER acknowledges a UK PPARC 
and a Greek IKY fellowship.
%
%
%
%

%


\section{FIGURE CAPTIONS}        

{\bf Figure 1} The observed flux calibrated spectra obtained with 
\cgs4 at \ukirt. The 14 SL sources are 
presented in RA order. The spectra of 3C~286 and the three 
serendipitous sources are given at the end. Note that there 
are only three SL sources where \Ha was not detected 
(the expected wavelength is indicated in 3C~216, 4C~29.45, and 
4C~56.27).

{\bf Figure 2} (a): The \Ha line  
versus continuum luminosity of the SL sample. (b): The EW 
versus the continuum luminosity. CDP, CD and LD objects are 
shown with cross, X and star symbols respectively.  
Small filled circle symbols show the \ukirt data. Upper limits 
are noted.

{\bf Figure 3} The \Ha EW versus  
\R , \ba, \d, and \th, (in a,b,c and d respectively) of the 22 SL 
objects. In these and all subsequent figures, circles designate 
the strong-EW objects (open and filled symbols are core and 
lobe dominated sources respectively). 
Note the inverted abscissa for the graphs versus \d , $R$, and 
\ba , chosen so that these can be easily compared with the 
correlation with \th .

{\bf Figure 4} The \Ha FWHM versus the orientation indicators as in Fig 3,
for the 19 objects with detected line emission.

{\bf Figure 5} (a): FWHM versus \th\ compared with models.
The dashed curve shows an axisymmetric component ($V=13,000 \sin \th $
\kms) and the solid curve gives its quadrature sum with 
an isotropic one ($V$=2,000 \kms, the dotted curve). 
(b): EW versus 
\th\ compared with models. Note the inverted EW axis indicating 
the brightness of the underlying continuum. The dot-dashed curve 
shows the expected effect of Doppler enhancement for a jet with \g =30.
The dashed curve shows the 
predicted effect for a flat accretion disc with a 
standard limb darkening, and the solid curve combines
disc and beamed component. The dotted curve combines
an isotropic and a beamed component.

{\bf Figure 6} Same as Fig 5, but with \th\, recalculated using \d\ bigger by 
factor three to show sensitivity to errors.

%
%
%
\begin{table*}
\centering
\caption{Names, redshift, SL speed and radio core dominance of the 
SL sample taken from VC94. The observed radio core flux and angular size 
at a given wavelength, and the X-ray flux at 1 keV are given together with 
the references.}
\begin{tabular}{llllrrrrrlrl}
\\[0.2cm]
\hline
\hline
\\[0.01cm]
\multicolumn{1}{c} {IAU }
&\multicolumn{1}{c} {Common}
&\multicolumn{1}{c} {Class}
&\multicolumn{1}{c} {z}
&\multicolumn{1}{c} {\ba$^1$}
&\multicolumn{1}{r} {log \R}
&\multicolumn{1}{c} {S$_\nu$ }
&\multicolumn{1}{c} {$\nu$}
&\multicolumn{1}{c} {$\theta_d$}
&\multicolumn{1}{l} {Ref.}
&\multicolumn{1}{c} {$F_X$}
&\multicolumn{1}{l} {Ref.}
\\
\multicolumn{1}{c} {name}
&\multicolumn{1}{c} {name}
&\multicolumn{1}{c} {}
&\multicolumn{1}{c} {}
&\multicolumn{1}{c} {c}
&\multicolumn{1}{r} {}
&\multicolumn{1}{c} {Jy}
&\multicolumn{1}{c} {GHz}
&\multicolumn{1}{c} {mas}
&\multicolumn{1}{l} {Radio}
&\multicolumn{1}{c} {$\mu$Jy}
&\multicolumn{1}{l} {X-ray}\\
\multicolumn{1}{c} {(1)}
&\multicolumn{1}{c} {(2)}
&\multicolumn{1}{c} {(3)}
&\multicolumn{1}{c} {(4)}
&\multicolumn{1}{c} {(5)}
&\multicolumn{1}{r} {(6)}
&\multicolumn{1}{c} {(7)}
&\multicolumn{1}{c} {(8)}
&\multicolumn{1}{c} {(9)}
&\multicolumn{1}{l} {(10)}
&\multicolumn{1}{c} {(11)}
&\multicolumn{1}{l} {(12)}\\
\vspace {0.1cm}\\
\hline
0007+106& III Zw 2$^2$&CD& 0.089&   1.25& $>$1.44& 1.540&  43&  0.075&B00
&2.74&B\\
0133+207& 3C 47   &LD& 0.425&   4.9& $-$1.30& 0.084&  10.7&  0.15&V93&  0.44&B\\
0430+052& 3C 120  &CD& 0.033&   5.4&  0.80& 3.900&   5.0&  0.40&P81& 10.00&BM\\
0552+398& DA 193  &CD& 2.365&   2.3&  0.60& 2.620&   8.4&  0.73&C90&  0.49&B\\
0850+581& 4C 58.17&CD& 1.322&   5.2&  0.27& 0.940&   5.0&  0.48&PR&  0.06&B\\
0906+430& 3C 216&CDP&   0.669&   5.1& $-$0.01& 0.880&   5.0&  0.10&PR&  0.11&B\\
0923+392& 4C 39.25&CD& 0.699&   5.3&  1.30& 0.200&  22.2&  0.05&A97&  0.58&B\\
1040+123& 3C 245  &CD& 1.029&   4.1&  0.00& 0.590&  10.7&  0.33&HR&  0.13&B\\
1156+295& 4C 29.45&CDP& 0.729&  34.8&  0.83& 1.400&  22.2&  0.12&H90&  0.18&B\\
1226+023& 3C 273  &CD& 0.158&  10.7&  0.90& 1.820&  22.0&  0.21&M00& 10.89&B\\
1253$-$055& 3C 279&CDP  & 0.538&  12.2&  1.10&12.310&  22.0&  0.08&W01&
0.96&B\\
1308+326& B2      &CDP& 0.996&  27.7&  1.60& 0.540&   5.0&  0.18&G93&  0.13&B\\
1618+177& 3C 334  &LD& 0.555&   2.5& $-$0.64& 0.086&  10.7&  0.20&H92&  0.20&B\\
1641+399& 3C 345  &CDP& 0.595&  12.4&  1.50& 2.750&  22.0&  0.13&RO&  0.57&B\\
1721+343& 4C 34.47&LD& 0.206&   3.2& $-$0.10& 0.109&  10.7&  0.24&H1&  2.37&B\\
1823+568& 4C 56.27&CDP& 0.664&   3.4&  2.30& 0.728&   5.0&  0.31&G94&
0.36&B95\\
1828+487& 3C 380  &CD& 0.691&  11.1&  0.70& 1.168&   5.0&  0.32&PW&  0.63&B\\
1830+285& 4C 28.45&LD& 0.594&   3.4& $-$0.28& 0.303&   5.0&  0.50&H2&  0.31&B\\
1845+797& 3C 390.3$^3$&LD& 0.057&   2.4& $-$1.00& 0.310&   5.0&  0.50&LO&  1.10&LO\\
1901+319& 3C 395  &CD& 0.635&  17.6&  0.50& 0.124&   4.8&  0.06&L99&... 
&...\\
1928+738& 4C 73.18&CD& 0.302&  9.4&  0.70& 2.110&   5.0&  0.49&PR&  1.05&B\\
2200+420& BL Lac  &CDP& 0.069&   5.0&  2.40& 1.600&   5.0&  0.35&M87&  0.82&M87\\
\hline
\multicolumn{12}{l}{1. Unless otherwise noted \ba is from VC94}\\
\multicolumn{12}{l}{2. \ba and \R\, from B00 and WB} \\
\multicolumn{12}{l}{3. \ba and \R\, from A96 and WB} \\
\\
\multicolumn{4}{l}{REFERENCES:}&\multicolumn{8}{l}{} \\
\multicolumn{4}{l}{A96: Alef \etal 1996}&\multicolumn{4}{l}{G94: Gabuzda \etal 1994} &\multicolumn{4}{l}{M00:
Mantovani \etal 2000} \\
\multicolumn{4}{l}{A97: Alberdi \etal 1997}&\multicolumn{4}{l}{H1: Hooimeyer \etal 1992}&\multicolumn{4}{l}{M87: Madau \etal 1987} \\
\multicolumn{4}{l}{B: Brinkmann \etal 1997}&\multicolumn{4}{l}{H2: Hooimeyer 
\etal 1992} &\multicolumn{4}{l}{P81: Pauliny-Toth \etal 1981} \\   
\multicolumn{4}{l}{B00: Brunthaler \etal 2000}&\multicolumn{4}{l}{H90: McHardy
\etal 1990}&\multicolumn{4}{l} {PR: Pearson \& Readhead 1988} \\
\multicolumn{4}{l}{B95: Brinkmann \etal 1995}&\multicolumn{4}{l}{H92: Hough
\etal 1992}&\multicolumn{4}{l}{PW: Polatidis \& Wilkinson 1998} \\  
\multicolumn{4}{l}{BM: Bloom \& Marscher 1991}&\multicolumn{4}{l} {HR: Hough
\& Readhead 1987}&\multicolumn{4}{l}{RO: Ros, Zensus \& Lobanov 2000}\\
\multicolumn{4}{l}{C90: Charlot 1990 }&\multicolumn{4}{l}{L99: Lara
\etal 1999}&\multicolumn{4}{l}{V93: Vermeulen \etal 1993}\\
\multicolumn{4}{l}{G93: Gabuzda \etal 1993}&\multicolumn{4}{l} {LO: Ledden
\& Odell 1985} &\multicolumn{4}{l}{W01: Wehrle \etal 2001}\\
\multicolumn{8}{l}{}&\multicolumn{4}{l}{WB: Wills \& Browne 1986}\\
\end{tabular}
\end{table*}
\begin{table*}
\centering
\caption{ The derived \d, \g\ and \th\ (in degrees) of the 
SL sources listed in increasing \th.}
\begin{tabular}{lrrrrrr}
\\[0.2cm]
\hline
\hline
\\[0.01cm]
\multicolumn{1}{l} {Name}
&\multicolumn{1}{r} {\d}
&\multicolumn{1}{r} {\g}
&\multicolumn{1}{r} {\th \deg}
\vspace {0.1cm}\\
\hline
 3C 279  &     73.45&     37.75&      0.25\\
 3C 216  &     42.58&     21.61&      0.32\\
 4C 29.45&      6.22&    100.55&      3.19\\
 1308+326&     11.59&     38.94&      3.52\\
 3C 395  &      5.07&     33.26&      6.00\\
 3C 345  &      8.45&     13.33&      6.32\\
 3C 380  &      6.22&     13.09&      7.86\\
 4C 73.18&      3.92&     13.29&     10.39\\
 3C 273  &      1.07&     54.94&     10.55\\
 4C 58.17&      5.42&      5.31&     10.62\\
 3C 120  &      5.32&      5.51&     10.82\\
 4C 56.27&      4.45&      3.64&     12.64\\
 BL Lac  &      4.43&      5.12&     12.92\\
 DA 193  &      3.76&      2.72&     13.99\\
 4C 39.25&      2.64&      6.83&     17.31\\
 III Zw 2&      2.43&      1.74&     21.11\\
 3C 47   &      0.50&     25.37&     22.75\\
 3C 245  &      1.73&      6.12&     23.41\\
 4C 28.45&      0.83&      8.02&     31.01\\
 4C 34.47&      0.19&     29.24&     35.05\\
 3C 334  &      0.40&      9.19&     42.68\\
 3C 390.3&      0.45&      7.85&     43.76\\        
\hline
\end{tabular}
\end{table*}
\begin{table*}
\caption{ The \ukirt\, observing log. Unless otherwise noted, the 
observations were made in $J$ band. 
}
\begin{tabular}{lcccl}
\\[0.2cm]
\hline
\hline
\\[0.01cm]
\multicolumn{1}{l} {Name}
&\multicolumn{1}{c} {Obs. date}
&\multicolumn{1}{c} {Seeing}
&\multicolumn{1}{c} {Exp. time}
&\multicolumn{1}{c} {Notes}\\
\multicolumn{2}{c} { }
&\multicolumn{1}{c} {arcsec}
&\multicolumn{1}{c} {min}
&\multicolumn{1}{c} { }
\vspace {0.1cm}\\
\hline
DA 193    &29/3/94& 1.16&46 &$K$, grating in first order\\ 
          &29/3/94& 1.35&72 &$H$, 600 \kms\\ 
4C 58.17  &29/3/94& 1.32&55 &$H$, 600 \kms\\
3C 216    &29/3/94& 1.31&27 &\\
4C 39.25  &27/3/94& 1.07&60 &2 objects in the slit\\
3C 245    &28/3/94& 1.10&80 &\\
4C 29.45  &28/3/94& 1.04&35  &\\
3C 279    &27/3/94& 1.23&88  &\\
1308+326  &27/3/94& 1.02&86  &\\
3C 286    &29/3/94& 1.06&75  &z=0.849\\  
3C 334    &27/3/94& 1.04&60  &\\
          &28/3/94& 1.30&85  &150 l/mm, R=1000\\
3C 345    &27/3/94& 1.08&60  &\\
4C 56.27  &29/3/94& 1.38&30  &\\
3C 380    &28/3/94& 1.27&40  &\\
4C 28.45  &28/3/94& 1.06&15  &\\
3C 395    &29/3/94& 1.15&30  &3 objects in the slit\\
\hline
\end{tabular}
\end{table*}
\begin{table*}
\centering
\caption{The observed \Ha\ flux (in 10$^{-14}$
\ergcms), rest frame EW and FWHM 
of the \ukirt\, \Ha spectra.}
\begin{tabular}{lrrr}
\\[0.2cm]
\hline
\hline
\\[0.01cm]
\multicolumn{1}{l} {Name}
&\multicolumn{1}{c} {Flux}
&\multicolumn{1}{c} {EW }
&\multicolumn{1}{c} {FWHM}\\
\multicolumn{1}{l} {}
&\multicolumn{1}{c} {}
&\multicolumn{1}{c} {\AA }
&\multicolumn{1}{c} {Km s$^{-1}$}

\vspace {0.1cm}\\
\hline
DA 193  & 6.00   &  240 & 2450 \\
4C 58.17  & 3.40   &  240 & 4100 \\
3C 216  &$<$0.05 &$<$10 &      \\
4C 39.25  & 8.60   &  295 & 4000 \\
3C 245  & 0.84   &  316 & 3250 \\
4C 29.45  &$<$0.06 &$<$8 &       \\
3C 279   & 1.40   &    8 & 1400 \\
 1308+326  & 0.90   &   30 & 2350 \\
3C 286  & 4.6    & 254  & 2600 \\ 
3C 334  &13.60   &  446 & 6150 \\
3C 345  & 3.25   &  186 & 3600 \\
4C 56.27 &$<$1.33 & $<$2 &      \\
3C 380  & 3.37   &  380 & 3500 \\
4C 28.45 & 6.67   &  354 & 4050 \\
3C 395  &28.60   &  158 & 2050 \\
\hline
\end{tabular}
\end{table*}
\begin{table*}
\centering
\caption
{The observed \Ha\ flux (in 10$^{-14}$
\ergcms), rest frame EW and FWHM 
and the references to the published \Ha spectra.}
\begin{tabular}{lrrrl}
\\[0.2cm]
\hline
\hline
\\[0.01cm]
\multicolumn{1}{l} {Name}
&\multicolumn{1}{c} {Flux}
&\multicolumn{1}{c} {EW }
&\multicolumn{1}{c} {FWHM}
&\multicolumn{1}{l} {Ref.}
\\
\multicolumn{1}{l} {}
&\multicolumn{1}{c} {}
&\multicolumn{1}{c} {\AA }
&\multicolumn{1}{c} {Km s$^{-1}$}
&\multicolumn{1}{l} {}\\
\vspace {0.1cm}\\
\hline
III Zw 2 &136&445&3600&N79,R85\\      
3C 47   &   10.5&     392&       8400&JB\\
3C 120  &   77.0 &    299&       1846&RB\\
3C 273  &   650.0 &    369&       3260&JB\\
4C 34.47&   65.0   &  589&       2300&S81, EH\\
3C 390.3&   80.6&     519&      11900&N82,L96,EH\\
4C 73.18&   38.0   &  365&       3100&JB,EH\\
BL Lac  &   28.7   &   7.3&       4000&C96\\
\hline
\\
\multicolumn{2}{l}{REFERENCES:}&\multicolumn{3}{l}{}\\
\multicolumn{5}{l}{C96: Corbett \etal 1996, EH: Eracleous \& Halpern 1994}\\
\multicolumn{5}{l}{JB: Jackson \& Browne 1991a, measured in scanned spectra}\\  
\multicolumn{5}{l}{L96: Lawrence \etal 1996, N82: Netzer 1982}\\
\multicolumn{5}{l}{N79: Neugebauer \etal 1979, R85: Rafanelli 1985}\\
\multicolumn{5}{l}{RB: Rokaki \& Boisson 1999,  
S81: Soifer \etal 1981}\\
\end{tabular}
\end{table*}
\begin{table*}
\centering
\caption{
The matrix of the
correlation coefficient $r_s$, and the two sided 
significance level of its deviation from zero, $p$, for
the sub-sample with detected \Ha\ lines (N=19) and for the 
subsample with strong lines (N=16). Correlations
with $p<0.02$ are shown in bold type. }  
\begin{tabular}{lrrrrrrrr}
\\[0.2cm]
\hline
\hline
\\[0.01cm]
\multicolumn{1}{l} {}  
&\multicolumn{2}{c} {FWHM}&
\multicolumn{2}{c}{EW}&  
\multicolumn{2}{c} {\lo}&
\multicolumn{2}{c} {\lha}\\ 
\multicolumn{1}{l} {}  
&\multicolumn{1}{c}{$r_S$}&  
\multicolumn{1}{c} {$p$}&
\multicolumn{1}{c}{$r_S$}&  
\multicolumn{1}{c} {$p$}&
\multicolumn{1}{c}{$r_S$}&  
\multicolumn{1}{c} {$p$}&
\multicolumn{1}{c}{$r_S$}&  
\multicolumn{1}{c} {$p$}\\
\hline 
\multicolumn{9}{c} {Strong- and weak-EW objects, N=19}\\
\hline
\\
EW   &    0.37&   0.115&&&&&&\\
\lo   &   -0.28&   0.244&   -0.42&   0.074&&&&\\
\lha  &    0.02&   0.949&    0.10&   0.687&    {\bf 0.71}&  {\bf 
0.001}&&\\
\\
\R   &   -0.38&   0.107&   {\bf -0.62}&  {\bf  0.005}&    0.08&   0.744&
-0.29&   0.228\\
\ba  &   -0.46&   0.047&   {\bf -0.59}&  {\bf 0.008}&    0.37&   0.115&   -0.01&   0.983\\
\d   &   -0.47&   0.043&   {\bf -0.78}& {\bf 7.5 10$^{-5}$}&    0.28&
0.241& -0.22&   0.363\\
\g   &   -0.32&   0.178&   -0.02&   0.930&    0.42&   0.075&    0.17&   0.477\\
\\
\th  &   {\bf 0.57}&   {\bf 0.011}&    {\bf 0.70}&  {\bf  0.001}&   -0.49&
0.033& -0.01&   0.955\\
\hline
\multicolumn{9}{c} {Strong-EW objects, N=16}\\
\hline
\\
EW   &    0.33&   0.214& &&&&&\\
\lo   &   -0.07&   0.805&   -0.52&   0.040&&&&  \\
\lha  &   -0.14&   0.615&   -0.34&   0.192& {\bf 0.96}& {\bf 5 10$^{-9}$}
&&\\
\\
\R   &   -0.41&   0.113&   -0.44&   0.092&    0.09&   0.747&    0.01&   0.976\\
\ba  &   -0.33&   0.215&   -0.52&   0.037&    0.25&   0.345&    0.15&   0.572\\
\d   &   -0.33&   0.207&   {\bf -0.74}& {\bf   0.001}&    0.17&   0.535&
0.01&   0.983\\
\g   &   -0.11&   0.674&    0.13&   0.623&    0.22&   0.412&    0.26&   0.333\\
\\
\th  &    0.46&   0.076&    {\bf 0.67}&   {\bf 0.005}&   -0.40&   0.122&
-0.28&   0.289\\
\hline
\end{tabular}
\end{table*}
%
%
%
%
\begin{figure*} 
\centering
\epsfxsize=17.cm
\epsfysize=24.cm
\caption{}
\hfill
\epsfbox{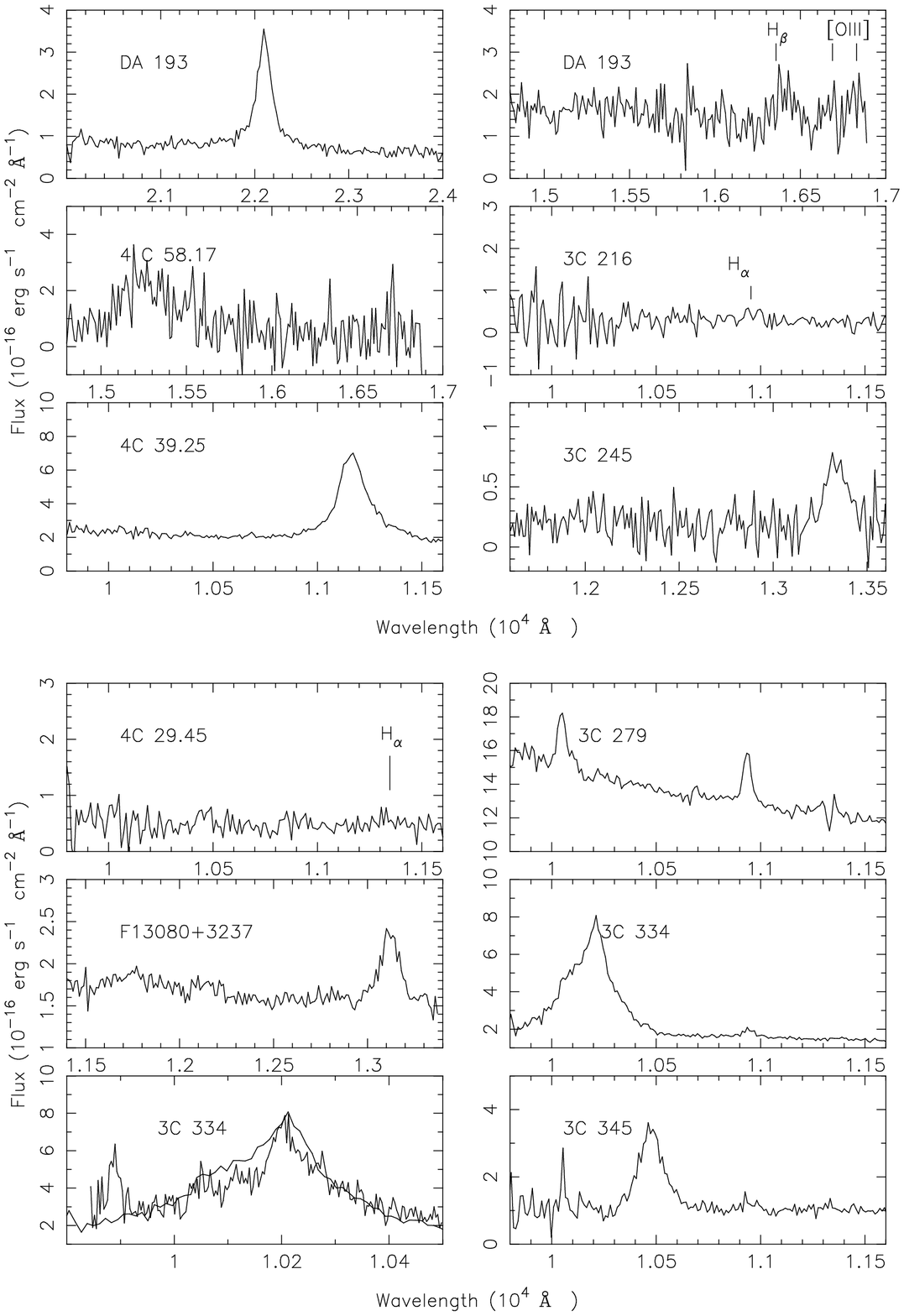}
\hfill
\epsfverbosetrue
\end{figure*}
\setcounter{figure}{0}
\begin{figure*}
\centering
\epsfxsize=17.cm
\epsfysize=24.cm
\caption{ - {\it continued}}
\hfill
\epsfbox{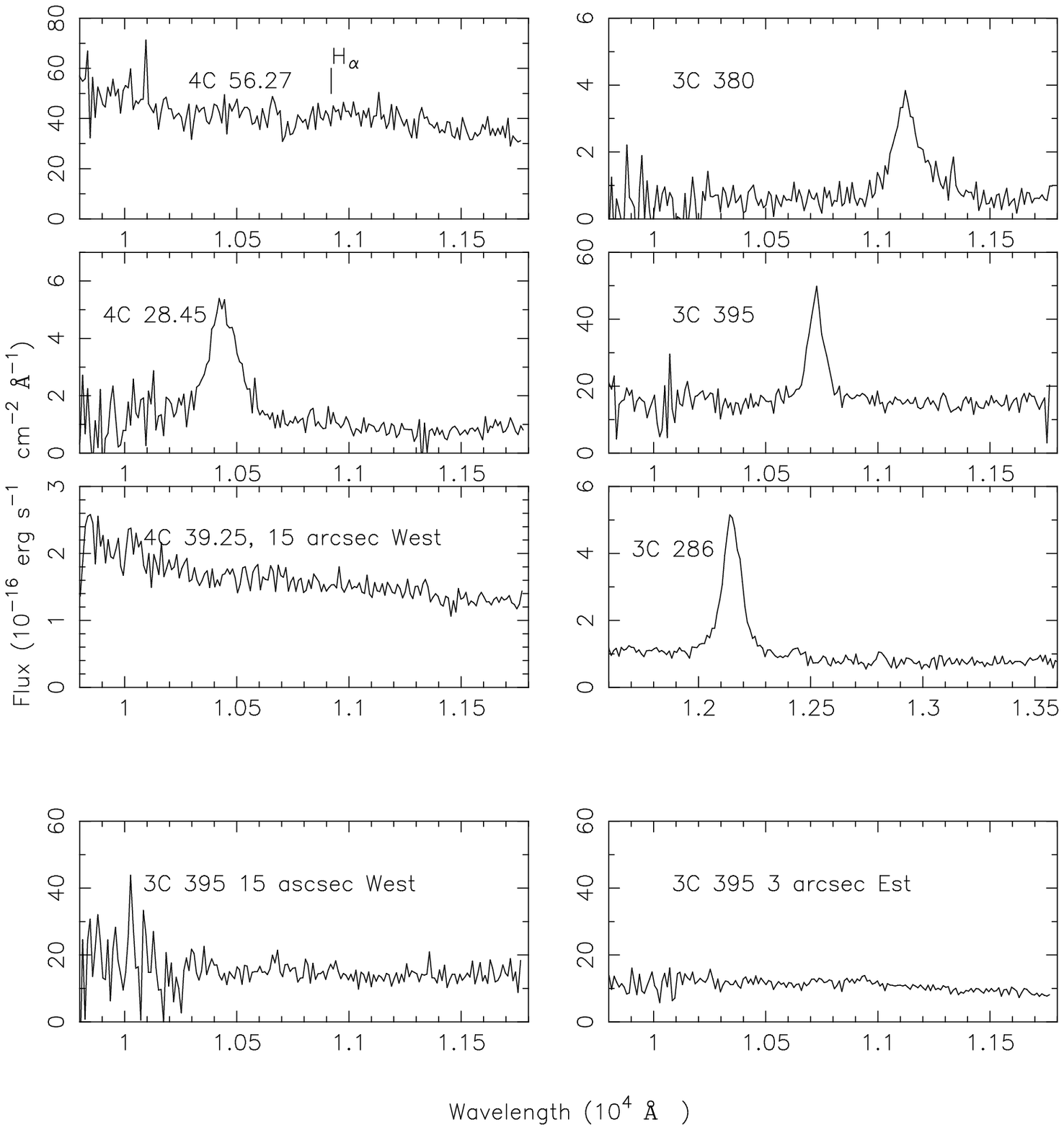}
\hfill
\epsfverbosetrue
\end{figure*}
\setcounter{figure}{1}
\begin{figure*}
\centering
\epsfxsize=17.cm
\epsfysize=24.cm
\caption{}
\hfill
\epsfbox{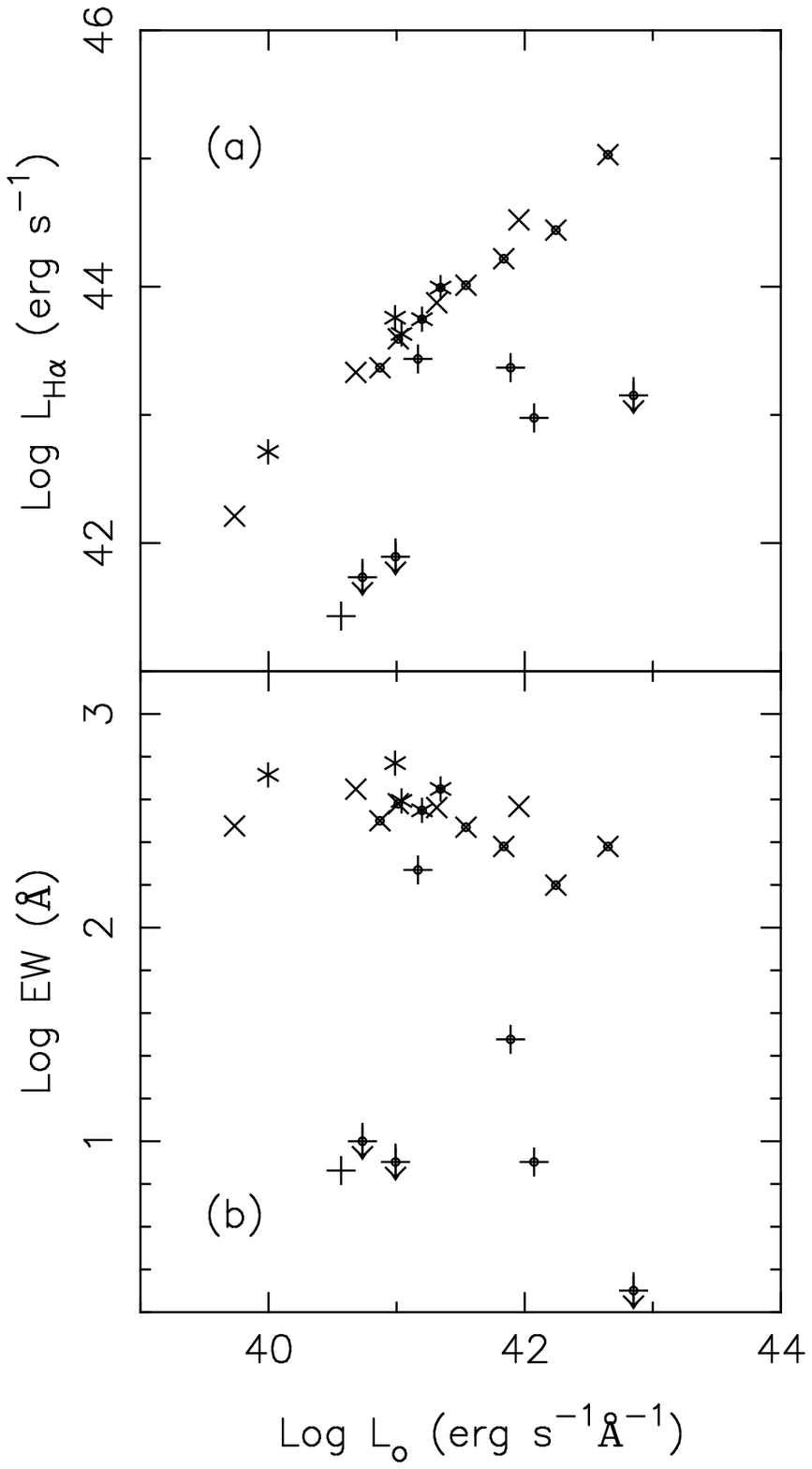}
\hfill
\epsfverbosetrue
\vspace{-3.6cm}
\end{figure*}
\begin{figure*}
\centering
\epsfxsize=17.cm
\epsfysize=24.cm
\caption{}
\hfill
\epsfbox{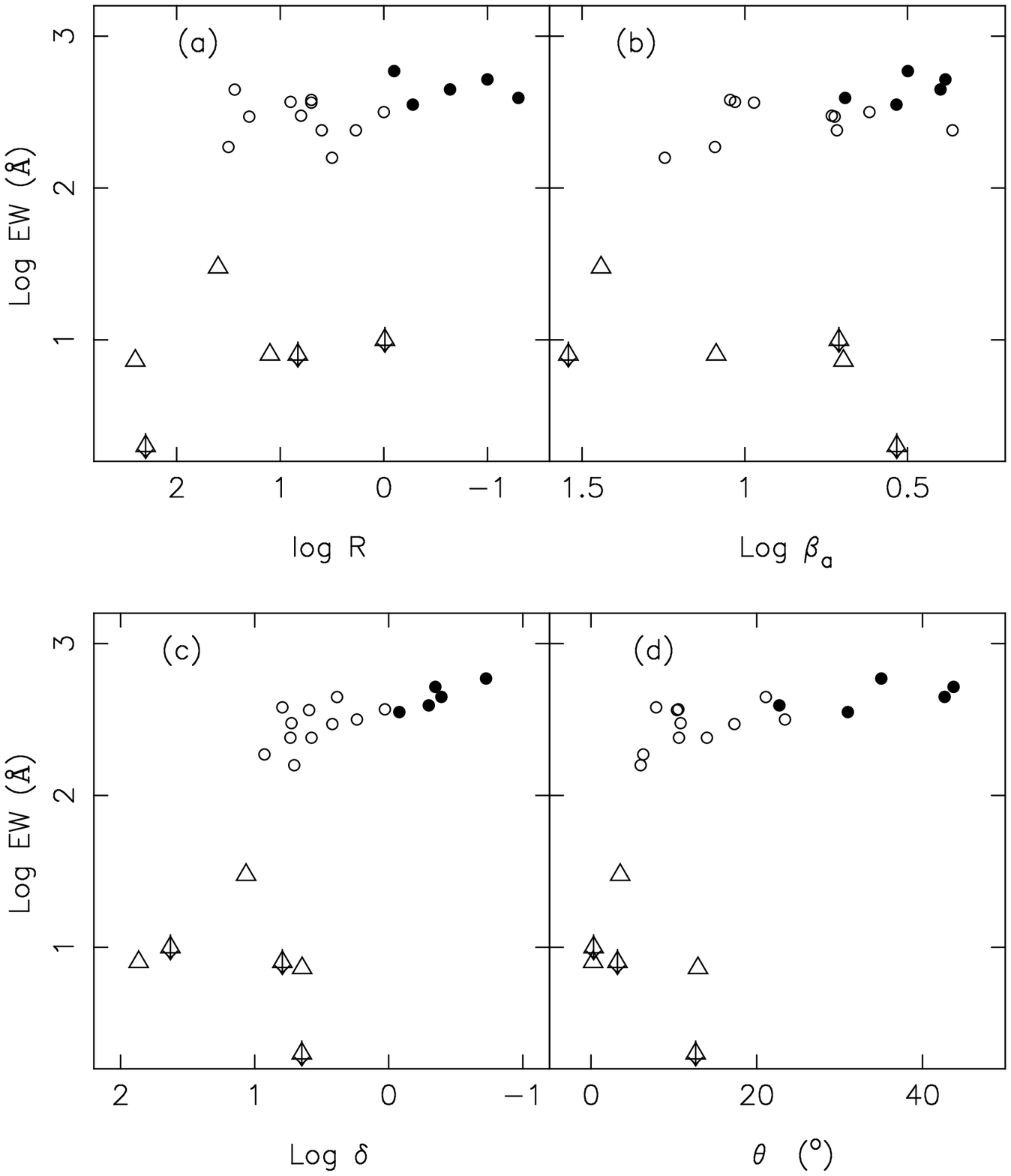}
\hfill
\epsfverbosetrue
\vspace{-3.6cm}
\end{figure*}
\begin{figure*}
\centering
\epsfxsize=17.cm
\epsfysize=24.cm
\caption{ }
\hfill
\epsfbox{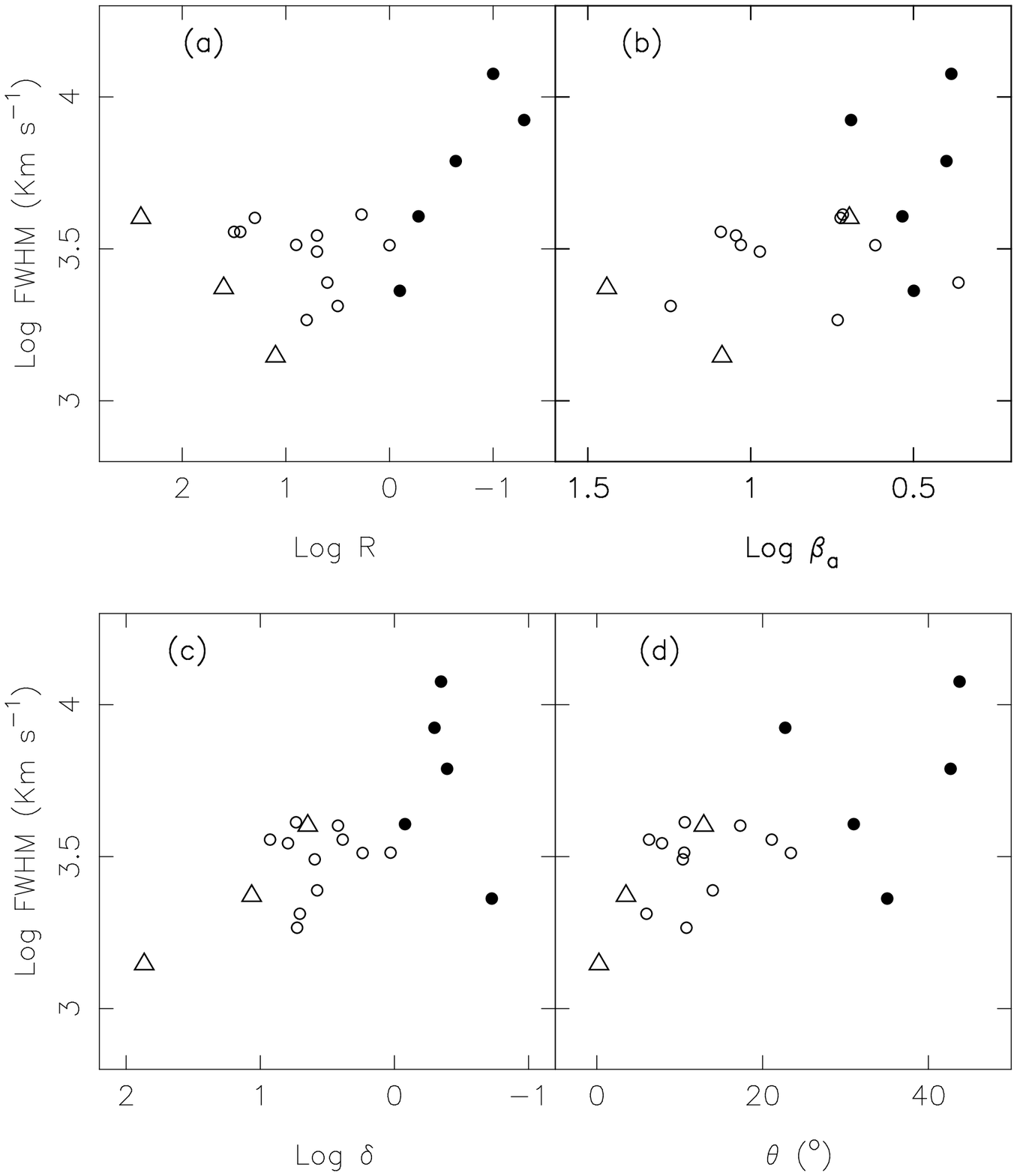}
\hfill
\epsfverbosetrue
\vspace{-3.6cm}
\end{figure*}
\begin{figure*}
\centering
\epsfxsize=17.cm
\epsfysize=24.cm
\caption{} 
\hfill
\epsfbox{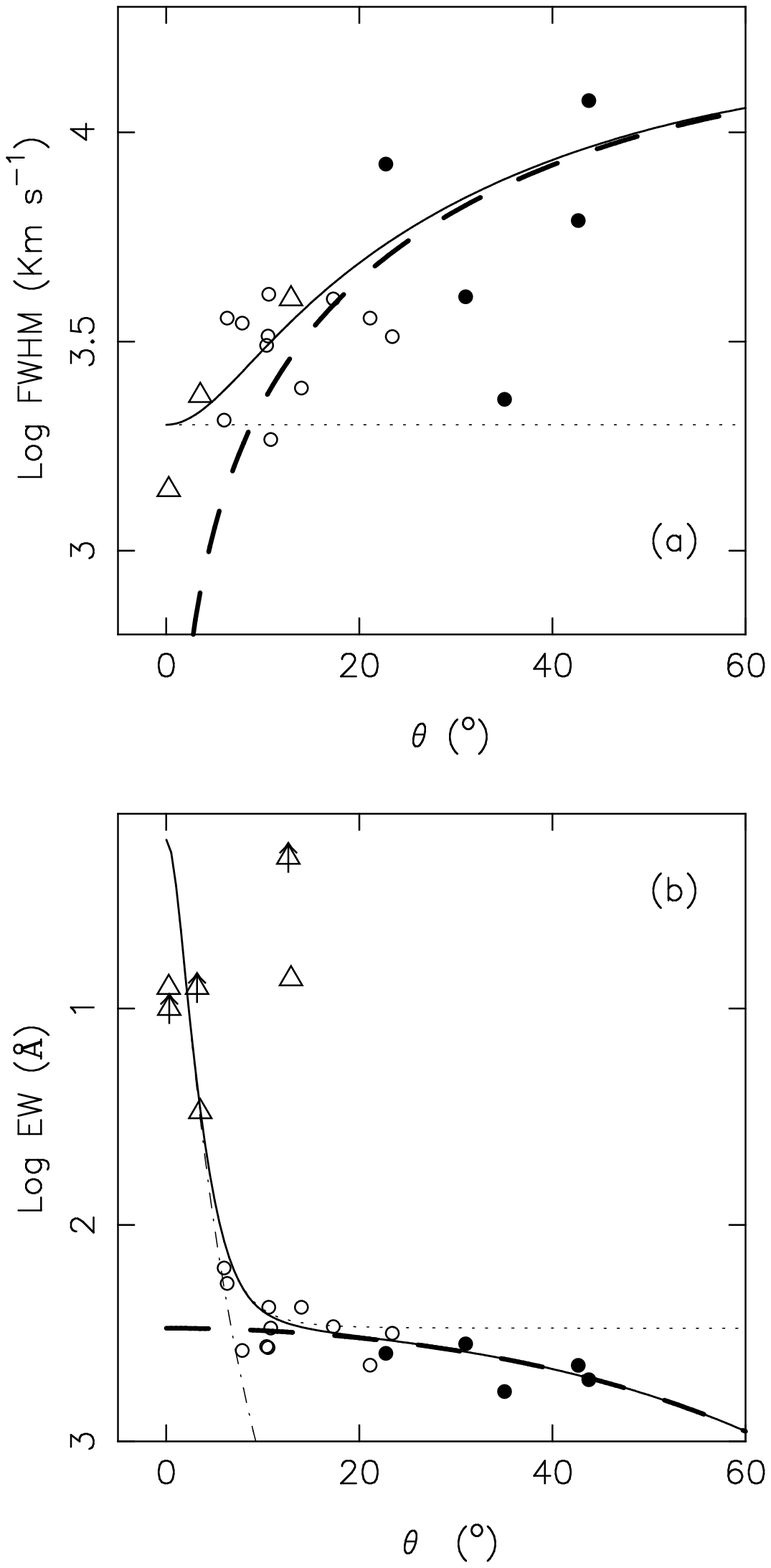}
\hfill
\epsfverbosetrue
\vspace{-3.6cm}
\end{figure*}
\begin{figure*}
\centering
\epsfxsize=17.cm
\epsfysize=24.cm
\caption{ } 
\hfill
\epsfbox{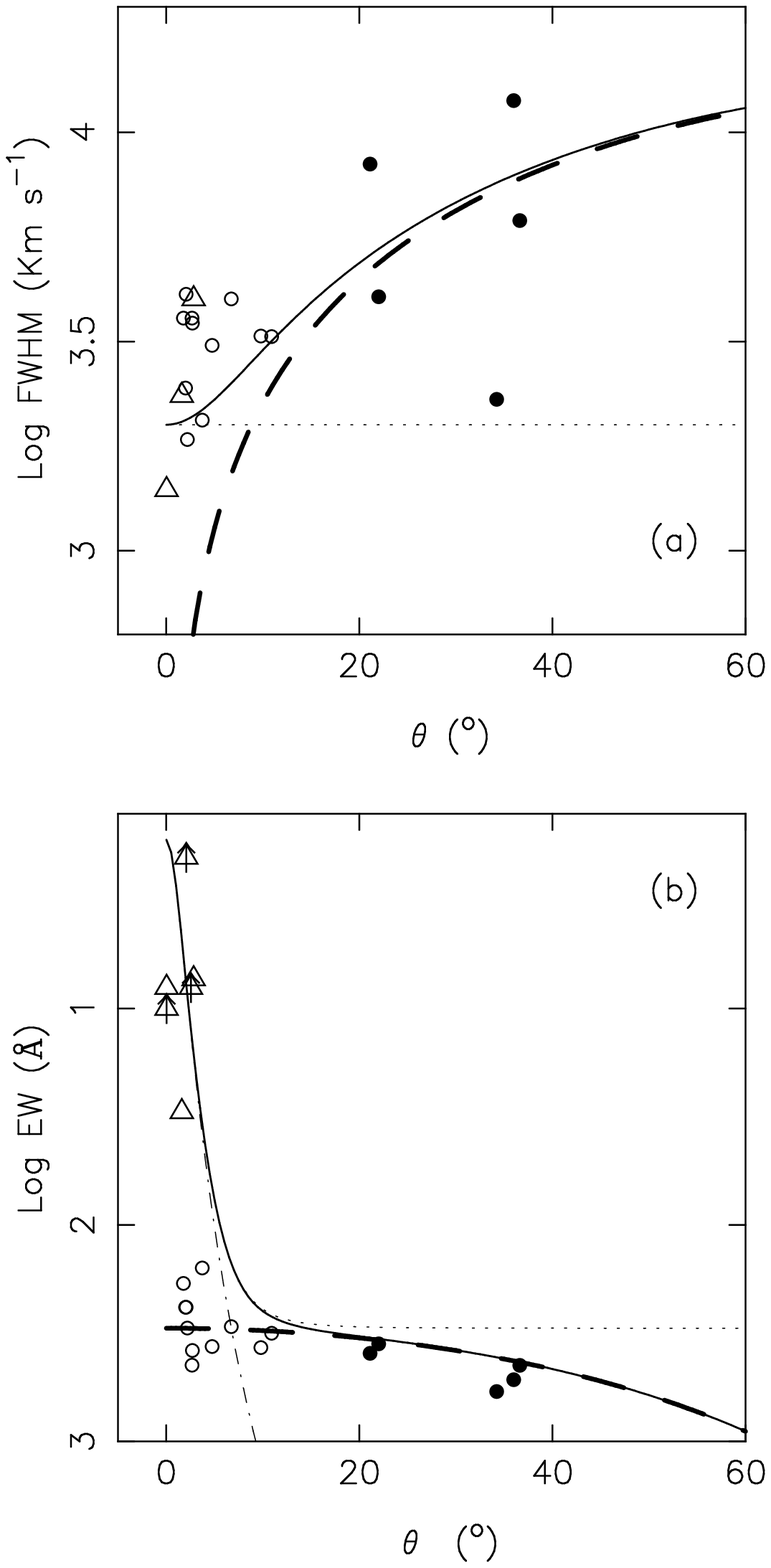}
\hfill
\epsfverbosetrue
\vspace{-3.6cm}
\end{figure*}
\end{document}